\documentclass[11pt]{article}
\usepackage{subfigure}
\usepackage{parskip}
 \expandafter\let\csname equation*\endcsname\relax
 \expandafter\let\csname endequation*\endcsname\relax
\usepackage{bm}
\usepackage{enumerate}
\usepackage[top=1in, bottom=1in, left=1in, right=1in]{geometry}
\usepackage{mathrsfs,color}
\usepackage{amsfonts}\usepackage{multirow}
\usepackage{amssymb}
\usepackage{amsmath}
\usepackage{float}
\usepackage{graphicx,afterpage}
\usepackage{epstopdf}
\usepackage[pdftex,colorlinks=true]{hyperref}
\usepackage[sort,compress]{cite}

\usepackage{natbib}

\usepackage{upgreek}
\usepackage[dvipsnames]{xcolor}
\usepackage{bm}
\usepackage{algorithm2e} 
\usepackage{hyperref}

\usepackage{multirow}

\usepackage{tikz}

\usepackage{xcolor}
\usepackage{tikz}
\hypersetup{
     colorlinks   = true,
     citecolor     = teal, % used blue to ease my editing
     linkcolor     = teal
}

\newcommand{\bx}{\mathbf{x}}
\newcommand{\by}{\mathbf{y}}

\newcommand{\bQ}{\mathbf{Q}}
\newcommand{\bM}{\mathbf{M}}
\newcommand{\bH}{\mathbf{H}}
\newcommand{\bR}{\mathbf{R}}

\newcommand{\bP}{\mathbf{P}}
\newcommand{\bK}{\mathbf{K}}

\newcommand{\bI}{\mathbf{I}}

\newcommand{\bldeta}{\boldsymbol{\eta}}
\newcommand{\beps}{\pmb{\varepsilon}}

\begin{document}
\begin{center}
{\bf\Large{Why we should condition denoising diffusion generative models on windows of past observations}}

\vspace{4mm}
Matthias Morzfeld$^1$, Daniel Hodyss$^2$

\vspace{2mm}
$^1$ Cecil H. and Ida M. Green Institute of Geophysics and Planetary Physics, Scripps Institution of Oceanography, University of California, San Diego, CA\\
$^2$ Remote Sensing Division, U.S. Naval Research Laboratory, Washington DC
\end{center}

\begin{center}
\emph{Abstract}
\end{center}
Data assimilation (DA) is, traditionally, a cycling process that relies on time-dependent priors to propagate information from past observations to future cycles.
Using denoising diffusion generative modeling for DA is challenging because standard approaches use a fixed training data set, which in turn leads to a static prior that ignores information from past observations.
Because past observations are ignored, DA systems with static priors lead to larger posterior errors than cycling DA systems.
Incorporating time-dependent priors into generative models, however, requires expensive and frequent retraining.
Motivated by linear systems theory -- where the dependence of a prediction of a Kalman filter on past observations decays exponentially -- we condition diffusion models on short windows of past observations. Specifically, we describe training procedures for two frameworks: a diffusion DA system predicting the current state given a set of past observations, and a diffusion ``direct observation prediction'' (DOP) system, predicting future observations given a set of past observations. Using a canonical linear system, we show that both systems can achieve the minimal posterior error characteristic of a fully-cycled DA\slash DOP system, without re-training, provided the time windows are long enough. The linear setup ensures analytical tractability, avoids confounding neural network training errors, and confirms that conditioning on windows of past observations is required for efficient and accurate diffusion-based DA or DOP.
\section{Introduction}
Numerical weather prediction (NWP) traditionally relies on a cycling data assimilation (DA) process over time windows: New observations, collected during the current time window, are used to update a forecast to an analysis, so that a new forecast can be generated for the next time window. 
The forecasts are generated by a ``physics-based'' model that approximates the time evolution of Earth's atmosphere, given a set of initial and boundary conditions.

The cycling process in DA ensures that information from past observations propagates, via the forecast model, to current and future analyses.
Mathematically, cycling DA requires time-dependent priors. In particular, these posterior distributions from the last cycle turn into prior distributions for the next cycle \citep[see, e.g.,][]{MH19}.
Moreover, the time-dependent priors are informed by \emph{all} past observations (due to cycling), which means that the priors have minimal variance.
By keeping deviations from the prior state minimal, traditional cycling DA systems can successfully apply linear tangent and adjoint models (in variational schemes) or linear update equations (ensemble Kalman filtering) to highly non-linear atmospheric dynamics.
Indeed, the various linear and Gaussian assumptions required by current ensemble DA (variational, ensemble or hybrid) may be one of the main reasons for frequent cycling in conventional DA systems.

Recent advances in machine learning (ML) and artificial intelligence (AI) challenge traditional DA in various ways, in particular because Ml\slash AI techniques are capable of fully nonlinear analyses that may not require nearly linear dynamics or nearly Gaussian priors.  
First, one may consider AI-tools (data-driven models) to replace the physics-based model to emulate and speed up the forecast model.
Three recent Google models, GraphCast \citep{graphcast}, neural-GCM \citep{neuralGCM} and GenCast \citep{gencast} are examples of AI forecasting tools that have reached or exceeded the skill of traditional systems, usually bench-marked by the forecast skill of the European Center for Medium-Range Weather Forecasts (ECMWF).
By now, AI forecasting tools are integrated into ECMWF's AI-FS \citep{AIFS} and are used operationally. 
We note that this integration happened at a rapid speed and in just a few years.

One may also consider replacing the entire DA system by an AI-tool.
Such DA systems are sometimes called ``end-to-end'' systems and the goal is to use an AI model that maps observations directly to future observations or to future atmospheric states.
At the time of writing, end-to-end systems are not as wide spread and not as well understood as the AI forecasting tools described above, but progress is rapid \citep{AIFSDOP}.

One candidate class of AI tools for ``pure'' AI-driven DA are denoising diffusion generative models.
Broadly, a diffusion model can learn an empirical distribution from training data and, once trained, produces new samples from that empirical distribution (see Section~\ref{sec:DiffusionReview} or, e.g., \cite{karras2022}).
For DA, a diffusion model should be trained to generate an ensemble of atmospheric states, given a set of observations.
Many researchers are currently building and testing diffusion DA systems of this type \citep[see, e.g.,][]{manshausen2024, Aardvark,rozet2023,simpleObLike2024,wang2026,BAO2024,bao2025nonlinear,bao2024ensemble,HM26}.
Diffusion DA systems, however, cannot easily propagate information from one cycle to the next.
The reason is that the training data essentially play the role of the prior:
If the diffusion is trained on the equivalent of one long ``free'' model run (usually on ERA5), then the prior is static, not time-dependent.
Consequently, such a diffusion DA system does not propagate information between successive assimilations.
With a less-informed, static prior (higher variance), diffusion DA systems have a lower accuracy than fully cycling DA systems \citep{HM26}.
The optimal accuracy of a cycling DA with time-dependent priors can be achieved by re-training, but re-training diffusion models is expensive.

In this paper, we demonstrate that diffusion DA systems benefit from conditioning on observations over a backward time-window during training.
We argue that this backward window of observations can be short, partially inspired by linear systems theory, where Kalman filter updates forget past observations exponentially fast \citep{ExponentiallyForgettingKFs}.
Indeed, we show for a canonical linear system that diffusion DA systems, conditioned on short backwards time windows, can achieve optimally small posterior error, matching  the accuracy of a fully cycled DA system.
Conditioning diffusion models on windows of past observations during training is thus essential for the accuracy of diffusion DA systems.
But conditioning the diffusion on windows of past observations also has computational advantages.
As we will show, the conditioning on windows of past observations emulates the effects of cycling priors, characteristic of traditional cycling DA systems, without the need for re-training the diffusion model.

% In a previous article \citep{HM26}, we showed that diffusion DA systems that do not cycle cannot match the optimal performance of an iteratively cycled filtering DA system.   
% Our goal in this article is to make use of this fact to show that diffusion DA systems configured without observational windows through time cannot be made optimally accurate.  
% It is the goal of this paper to explore this fact and illustrate clearly why non-cycled DA, for which typical diffusion DA systems are good examples, must use appropriately configured observation windows through time.

The rest of this paper is organized as follows.
Section~\ref{sec:Background} reviews the necessary background for this paper: Traditional DA via Kalman filters (Section~\ref{sec:TraditionalDAReview}), direct observation prediction (DOP) via Kalman filters (Section~\ref{sec:DOPReview}), and denoising diffusion generative modeling (Section~\ref{sec:DiffusionReview}).
We describe in particular how to derive exact denoisers for reverse processes for Gaussian problems to avoid training neural networks.
Section~\ref{sec:SourcesOfErrors} explains that three sources of error interact in a diffusion DA system: Time discretization errors of the reverse process, Monte Carlo errors in the estimates of the loss functions, and approximation errors of neural networks that parameterize the denoiser.
Section~\ref{sec:MemorySection} describes how to define training data sets for diffusion DA systems (Section~\ref{sec:DiffDAMemory}) and diffusion DOP systems (Section~\ref{sec:DiffDOPMemory}) that leverage backward time windows of observations.
We work out a case study with a canonical linear system in Section~\ref{sec:NumericalIllustration} to illustrate our ideas and to show that diffusion DA\slash DOP systems conditioned on short backward time windows can be optimally accurate and indeed as accurate as a fully cycling DA\slash DOP system.
Throughout, we focus on the idealized scenario in which the approximation errors discussed in Section~\ref{sec:SourcesOfErrors} are small, but we also show how neural network approximations within the diffusion model introduce additional approximation errors.
We conclude with a summary of our findings and a discussion of the implications of our study in Section~\ref{sec:SummaryAndDiscussion}.

\section{Background: Data assimilation, direct observation prediction and diffusion modeling}
\label{sec:Background}
We briefly review the basics of data assimilation (DA), direct observation prediction (DOP) and diffusion modeling.
We focus on linear\slash Gaussian systems because subsequent analysis and numerical experiments are carried out in a linear Gaussian setup (and we explain why further below).

\subsection{Data assimilation}
\label{sec:TraditionalDAReview}
The goal in traditional data assimilation (DA) is to update the state of a dynamical system iteratively with information from observations.  We focus here on relating filtering DA to diffusion modeling and we leave the comparison to a smoother for subsequent work.
From the perspective of filtering DA, we denote the state at discrete time $k\geq 0$ by the $n_x$ dimensional vector $\bx_k$ and the observation at time $k$ by the $n_y$ dimensional vector $\by_k$.  We thus compute or approximate the Bayesian posterior distribution of the state at time $k$, given the observations $\by_{1:k} = \{\by_1,\by_2,\dots,\by_{k}\}$:
\begin{equation}
\label{eq:Posterior}
    p(\bx_k\vert \by_{1:k}) = p(\by_k\vert \bx_k) p(\bx_k\vert \by_{1:k-1}).
\end{equation}
Here, the term $p(\by_k\vert \bx_k)$ is the likelihood, defined by the observation model and $p(\bx_k\vert \by_{1:k-1})$ is a time-dependent (cycling) prior \citep{HM26}.  
Note that the time-dependent prior $p(\bx_k\vert \by_{1:k-1})$ includes \emph{all} past observations, $\by_{1:k-1}$.
For that reason, the time-dependent prior has a smaller variance than, e.g., a climatological prior, $p(\bx_k)$.  
Hence, the correction required to transition from the time-dependent prior to the posterior is smaller than the correction required to transition from the climatological prior to the posterior.
 
For a linear\slash Gaussian system the dynamics and observations are linear functions of the state, perturbed by mean-zero Gaussian noise:
\begin{eqnarray}
    \label{eq:Model}
    \bx_k  =& \bM \bx_{k-1} + \beps_{k-1},   & \beps_{k-1}\sim\mathcal{N}(0,\bQ), \text{ iid},\\
    \label{eq:Observations}
    \by_k  =& \bH \bx_k + \bldeta_k,  & \bldeta_k\sim\mathcal{N}(0,\bR), \text{ iid},
\end{eqnarray}
where $\bM$ and $\bH$ are $n_x\times n_x$ and $n_y\times n_x$ matrices;
$\beps_k$ and $\bldeta_k$ are independently and identically distributed (iid) mean zero Gaussian random variables with symmetric positive definite (SPD) covariance matrices $\bQ$ and $\bR$ respectively ($\bQ$ is $n_x\times n_x$ and $\bR$  $n_y\times n_y$).
Under linear and Gaussian assumptions, the state conditioned on the observations is Gaussian distributed and the Kalman filter \citep{K60} is an efficient means to iteratively compute the mean and covariance of the Bayesian posterior distribution in~\eqref{eq:Posterior}, defined by the model and observation equations~\eqref{eq:Model} and~\eqref{eq:Observations}.

Specifically, suppose the posterior probability density function (pdf) at time $k-1$, $p(\bx_{k-1}\vert \by_{1:k-1})$, is given by the Gaussian
\begin{equation}
    \bx_{k-1}\vert \by_{1:k-1} \sim \mathcal{N}(\boldsymbol{\mu}_{k-1},\bP_{k-1}).
\end{equation}
To compute the pdf at time $k$, we first compute the ``forecast'' mean and covariance
\begin{equation}
    \boldsymbol{\mu}_\text{f} = \bM \boldsymbol{\mu}_{k-1},\quad \bP_\text{f} = \bM \bP_{k-1} \bM^\top+\bQ,
\end{equation}
and then perform an analysis step which accounts for the current observations $\by_k$:
\begin{eqnarray}
    \bK & = & \bP_\text{f}\bH^\top (\bH\bP_\text{f}\bH^\top+\bR)^{-1},\\
    \boldsymbol{\mu}_k & = & \boldsymbol{\mu}_\text{f}+\bK(\by_k-\bH\boldsymbol{\mu}_\text{f}),\\
    \bP_k & = & (\bI - \bK\bH)\bP_\text{f}.
\end{eqnarray}
The result of the analysis is the updated posterior distribution $\bx_{k}\vert \by_{1:k} \sim \mathcal{N}(\boldsymbol{\mu}_{k},\bP_{k})$.

The ensemble Kalman filter (EnKF) \citep[see, e.g.,][]{E94,E09Book,TetAl03,OEtAl04,BvLE98,EAKF} can be used when the model and observation equations are nonlinear. 
The EnKF has proven to be efficient and effective in numerical weather prediction, provided localization and inflation are implemented to keep the ensemble size moderate \citep[see, e.g.,][]{MH23,HM23, A12,A07,A03,HKS07,F15,N15,Menetrier15} and provided that the DA problem is mildly nonlinear \citep{MH19}.

In the presence of stronger nonlinearity, ``hybrid'' variational methods \citep[see, e.g.,][]{HS00, L03, ZZH09, BMC13, KRBMB13, PZ15}, that combine the Monte Carlo approach of an EnKF with optimization characteristic of a variational DA system \citep[see, e.g.,][]{T87}, have proven to be the most accurate DA systems over the past few decades.

\subsection{Direct observation prediction}
\label{sec:DOPReview}
Direct observation prediction (DOP) makes predictions of future observations based on past observations.
A natural target distribution is, thus,
\begin{equation}
\label{eq:DOP_pdf}
    p(\by_{k+1} \vert \by_{1:k}),
\end{equation}
but, most DOP systems use a sliding window of observations \citep{AIFSDOP,AIDOP,AIDOPUpdate,GraphDOP,GraphDOPInsights,CoupledGraphDOP} and the targeted pdf is
\begin{equation}
    p(\by_{k+1} \vert \by_{k-m:k}),
\end{equation}
for $0<m<k$. 
Here, the integer $m$ defines the sliding window, which, at the time of writing and for the latest European Center for Medium-Range Weather Forecasts (ECMWF) DOP system is two days \citep{AIFSDOP}.

In the linear\slash Gaussian setup in equations~\eqref{eq:Model} and~\eqref{eq:Observations}, the pdf~\eqref{eq:DOP_pdf}, relevant to DOP, is a Gaussian and we can compute the mean and covariance as follows.
At time $k$, the posterior distribution over the system state $\bx_k$ is Gaussian with mean $\boldsymbol{\mu}_k$ and covariance matrix $\bP_k$ (computed via the Kalman filter above).
The predicted observation is, thus
\begin{equation}
    \by_{k+1}\vert \by_{1:k} = \bH\bM\cdot (\bx_k\vert \by_{1:k}) +\boldsymbol{\eta},
\end{equation}
where $\boldsymbol{\eta}$ is a mean zero Gaussian with covariance matrix $\bR$,
so that 
\begin{equation}
    \label{eq:DOPTargetPDF}
    \by_{k+1} \vert \by_{1:k} \sim\mathcal{N}( \bH\bM\boldsymbol{\mu}_k,\bH\bM\bP_k\bM^\top \bH^\top+\bH\bQ\bH^\top+\bR).
\end{equation}
DOP systems generally avoid using the model $\bM$ or the observation operator $\bH$ and we explain below how this can be achieved via denoising diffusion generative modeling.
We use~\eqref{eq:DOPTargetPDF} as a reference to assess the accuracy of a DOP system in a linear\slash Gaussian setting.

We acknowledge that our treatment of DOP is simplified compared to what has been implemented in practice.
Modern DOP systems \citep{AIFSDOP} do not only forecast the next observations, but in fact forecast the full state.
In practice, full state prediction from observations alone is done by ``warm-up'' cycles in which the DOP system is producing and processing full state estimates \citep{AIFSDOP}.
This warm-up process is more difficult to study with mathematical rigor and we view our work below as a first step towards a mathematical understanding and theory for DOP systems that make full-state predictions.

\subsection{Denoising diffusion generative modeling}
\label{sec:DiffusionReview}
We explain the basics of a denoising diffusion generative model and distinguish between (unconditional) diffusion models and conditional diffusion models.
For clarity and ease of presentation, we consider only the scalar case, but the theory naturally extends to vectors (see, e.g., \cite{HM26} for examples).

\subsubsection{Unconditional denoising diffusion generative modeling}
Given a set of $n_s$ data samples $\{s_i\}$, $i=1,\dots,n_s$, from a distribution $p(s)$,
we want to generate more samples that also have the distribution $p(s)$.  
Recall that the target pdf $p(s)$ is not known analytically, but rather is only known through a long training set assumed to be drawn from this distribution.  In NWP, this type of training data is typically obtained from a long-time series of previous DA products, observations, or a very long model run.  
\cite{HM26} consider several training sets that lead to different types of diffusion models.  
We focus on the ``standard'' technique, referred to as the ``climatological prior,'' in \cite{HM26}, which is consistent with a fixed, long time series (e.g., ERA5) .  

To generate samples, we define a forward process~$u$ and a reverse process $v$, which satisfy stochastic differential equations (SDE).
We follow \cite{karras2022} and consider a variance exploding forward process $u$ 
\begin{equation}
    \label{eq:ForwardProcess}
    \text{d}u = \sqrt{2t_\text{d}}\text{d}W_\text{f},
\end{equation}
where $t_\text{d}$ is the diffusion time (not to be confused with ``physical'' time) and where $W_\text{f}$ is a Wiener process with the properties that $W_\text{f}$ (i) starts at zero, $W_\text{f}(0)=0$; (ii) $W_\text{f}$ has independent Gaussian increments $\Delta W_\text{f} = W_\text{f}(t_\text{d}+\Delta t_\text{d}) - W_\text{f}(t_\text{d})\sim \mathcal{N}(0,\Delta t_\text{d})$; and (iii) $W_\text{f}$ is almost surely continuous.
We initialize the forward process with a sample $u(0) = s_i$ and simulate the forward process up to time $T\gg 0$.
Note that the forward process amounts to sequentially adding Gaussian noise to the sample $s_i$.
The corresponding reverse process $v$ is the SDE
\begin{equation}
    \label{eq:ReverseProcess}
	\text{d}v = -2 \frac{E[u\vert v]-v}{t_\text{d}}
	\text{d}t_\text{d} + \sqrt{2t_d}\text{d}W_\text{r},
\end{equation}
where $W_\text{r}$ is also a Wiener process and where the conditional expectation is also called the ``denoiser:'' 
\begin{equation}
\label{eq:UnconditionalDenoiser}
    D(v,t_\text{d}) = E[u\vert v] = \int_{-\infty}^{\infty} u \, p(u\vert v)\text{d}u.
\end{equation}
To draw samples from $p(s)$, we initialize the reverse process at a large time $T\gg0$ and then integrate backwards in time to $t_\text{d}=0$.
At time $t_\text{d}=0$, the distribution of $v_0$ is $p(\cdot)$.

\subsubsection{Computing the denoiser}
\label{sec:ApproximateAndExactDenoisers}
The reverse process~\eqref{eq:ReverseProcess} requires that we compute the denoiser~\eqref{eq:UnconditionalDenoiser}. 
The common strategy is to parameterize the denoiser by a deep neural network (NN) and train the network on the loss function
\begin{equation}
    \label{eq:UnconditionalDenoiserLoss}
    \mathcal{L}(\boldsymbol{\theta}) =  E_{s \sim p_s(\cdot)}E_{n\sim \mathcal{N}(0,t_\text{d}^2)}\left\|s-D_\text{NN}(\boldsymbol{\theta}; v=s+n,t_\text{d})\right\|_2^2,
\end{equation}
whose minimizer is the conditional mean~\eqref{eq:UnconditionalDenoiser}, and where we use $\boldsymbol{\theta}$ to denote the network parameters, i.e., the weights and biases of the NN that define the denoiser.  
Typically, the denoiser is trained once and the resulting diffusion model is deployed.
The once-and-for-all training, however, implies a static prior, which is problematic for DA \citep{HM26}.

In simple cases, we can compute the denoiser analytically, 
\emph{without} using NNs.
If the samples $s_i$ are generated by a Gaussian with mean $\mu$ and standard deviation $\sigma$, then Bayes' rule implies that
\begin{equation}
    p(u\vert v) \propto p(u) p(v\vert u),
\end{equation}
where $v$ is the reverse process at time $t_\text{d}$, $p(u)$ is the Gaussian target distribution and where
\begin{equation}
    p(v\vert u) \propto \exp\left(-\frac{1}{2} \left(\frac{u-v}{t_\text{d}}\right)^2\right),
\end{equation}
due to our simple choice of the forward process.
Thus, $p(u\vert v)$ is the Gaussian
\begin{equation}
    p(u\vert v) \propto \exp\left(-\frac{1}{2} \left(\frac{u-m}{\sigma}\right)^2\right) \exp\left(-\frac{1}{2} \left(\frac{u-v}{t_\text{d}}\right)^2\right),
\end{equation}
whose mean is
\begin{equation}
    \label{eq:ExactGaussianDenoiser}
    D(v,t_\text{d}) = E[u\vert v] = \mu + \frac{\sigma^2}{\sigma^2 +t_\text{d}^2}(v-\mu).
\end{equation}
We note that the above formula generalizes easily to multivariate Gaussians (see \cite{HM26} for examples).
% \begin{equation}
%     D(\mathbf{v},t) = E[\mathbf{x}\vert \mathbf{v}_t] = \mathbf{m} + \bP(\bP + t^2\bI)^{-1}(\mathbf{v}_t-\mathbf{m}).
% \end{equation}

\subsubsection{Conditional denoising diffusion generative modeling}
In addition to sampling an empirical distribution defined by the training samples, i.e., sampling $p(s)$, it is also possible to sample conditional distributions with extra inputs, say $p(s\vert y)$, where $y$ are additional inputs.
For example, one can use denoising diffusion models to generate images conditioned on a text prompt: ``\emph{Generate an image of a surfing elephant}''. In this context, $y$ are the text prompts and the training data set requires a set of images (samples $s_i$) with accompanying text prompts, i.e., $\{s_i,y_i\}$ pairs.
The construction of the denoising diffusion is similar to what we described above, but the denoiser in~\eqref{eq:ReverseProcess} now requires $y$ as additional inputs:
\begin{equation}
    D(v,y,t_\text{d}) = E[u\vert v,y] = \int_{-\infty}^{\infty} u\, p(u\vert v,y)\text{d}u.
\end{equation}
The loss function that we use to train an NN to calculate the denoiser thus becomes
\begin{equation}
    \label{eq:ConditionalDenoiserLoss}
    \mathcal{L}(\boldsymbol{\theta}) =  E_{s \sim p(s, y)}E_{n\sim \mathcal{N}(0,t_\text{d}^2)}\left\|s-D_\text{NN}(\boldsymbol{\theta}; v=s+n,y,t_\text{d})\right\|_2^2.
\end{equation}

If $p(s\vert y)$ is Gaussian, we can compute its mean $\mu$ and variance $\sigma^2$ and define an ``exact'' Gaussian denoiser as in~\eqref{eq:ExactGaussianDenoiser}, using Bayes' rule for the conditional density
\begin{equation}
    p(s\vert y,v)\propto p(s\vert y) p(v\vert s),
\end{equation}
noting that $p(v\vert s,y) = p(v\vert s)$, due to the simple construction of the forward process~\eqref{eq:ForwardProcess}.

The differences between conditional and unconditional denoising diffusion generative models are stark.
In the context of DA and NWP, an unconditional diffusion model samples a typical atmospheric state.
A conditional diffusion can take in extra inputs in the form of past states and\slash or past observations to generate atmospheric states that are compatible with atmospheric conditions \citep{HM26,gencast}.
Conditional denoising diffusion models are thus scientifically more useful than unconditional denoising diffusion models, which explains the increase of research activity in this field \citep[see, e.g.,][]{manshausen2024, Aardvark,rozet2023,simpleObLike2024,wang2026}. 

The AI literature agrees that conditional diffusion models are computationally more expensive to build and to train than unconditional diffusion models.  
As a consequence, many researchers train an unconditional diffusion and then ``guide'' the unconditional diffusion towards the observations \cite[see, e.g.,][]{rozet2023,simpleObLike2024,wang2026,manshausen2024}.  
We will not consider guided diffusion models in this paper and we will explain why after we discuss the importance of training conditional diffusion models on windows of past observations.

\section{Sources of errors in denoising diffusion generative models}
\label{sec:SourcesOfErrors}
Denoising diffusion generative models are defined by continuous forward and reverse processes (equations~\eqref{eq:ForwardProcess} and~\eqref{eq:ReverseProcess}) and the loss functions associated with the denoisers (equations~\eqref{eq:UnconditionalDenoiserLoss} and~\eqref{eq:ConditionalDenoiserLoss}) are formulated in terms of expectations over random variables.
In practice, we discretize the forward and reverse processes and use Monte Carlo approximations for the expected values in the loss functions~\eqref{eq:UnconditionalDenoiserLoss} or~\eqref{eq:ConditionalDenoiserLoss}.
The training data comes into play in the Monte Carlo approximations, where the expected values in~\eqref{eq:UnconditionalDenoiserLoss} or~\eqref{eq:ConditionalDenoiserLoss} become sums (averages) over training samples $s_i$ and noising steps ($n$).

Practical denoising diffusion systems are therefore characterized by three sources of discretization errors.
\begin{enumerate}
    \item \emph{Time discretization errors} in the forward and reverse processes:
    Keeping these errors small requires high-order SDE solvers and\slash or small time steps when simulating the reverse process for sample generation.
    \item \emph{Monte Carlo errors} in the approximation of the loss functions~\eqref{eq:UnconditionalDenoiserLoss} or~\eqref{eq:ConditionalDenoiserLoss}.
    Keeping these errors small requires a large number of training samples and a large number of noising steps, i.e., a large training data set.
    \item Errors in the \emph{NN approximations} of the denoiser. 
    Keeping these errors small requires a careful design and training of the NN.
\end{enumerate}

Below, we build conditional denoising diffusion models for scalar, linear and Gaussian DA and DOP systems.
The linear\slash Gaussian setup allows us to separate the three sources of error in denoising diffusion generative models.
First, we can avoid Monte Carlo errors and errors due to NN approximations of the denoiser by computing analytical denoisers for Gaussian distributions. 
With Monte Carlo and NN approximation errors out of the picture, we can investigate the effects of time discretization errors, but our numerical experiments below suggest that these errors are negligible.

The analytical denoisers effectively mimic a well-designed and well-trained denoising diffusion generative model, with access to a large training data set.
These assumptions are not unreasonable given that (i) decades of atmospheric data are available for training; and (ii) the success of generative AI suggests that sufficiently accurate NN denoisers are within reach, provided one has access to sufficiently large computing resources.

The linear\slash Gaussian setup thus indicates best-case performance scenarios for denoising diffusion generative models for DA and DOP and it allows us to highlight the importance of incorporating time windows of observations into diffusion DA and DOP systems, which us our main point.

\section{Diffusion models with observational memory for DA and DOP}
\label{sec:MemorySection}
We describe how to design the training data for diffusion for DA or DOP systems.
Our goal is to convincingly argue that diffusion DA\slash DOP should have an ``observational memory,'' i.e., the conditioning should include \emph{several} past observations \citep{HM26}, not just the current observation, as in some current diffusion DA systems \citep[e.g.,][]{manshausen2024}. 
%[Are you sure these guys use obs at a single time?  I'm not sure Allen et al does.  Manshausen might.  They admit the obs are random in time.  Then they do the really strange thing of interpolating them to the ``hour", but do they actually say they assimilate only one hour at a time?  I couldn't find that.  At least in operational meteorology, I'm not certain that filtering actually exists.  As far as I know all operational NWP uses windows.  The only people I'm sure of who use filters are academic scientists (like us!) writing papers.  That's why we should carefully verify who actually might use a filter.]
% I checked! Good catch, these other two really do use observations distributed in time

\subsection{Diffusion models with observational memory for DA}
\label{sec:DiffDAMemory}
As stated before, the goal in DA is to approximate the posterior distribution of the current state, conditioned on \emph{all} past observations,  $p(\bx_k\vert \by_{1:k})$ (see~\eqref{eq:Posterior}).
For a \emph{linear} system, however, the Bayesian posterior distribution~\eqref{eq:Posterior} is well approximated by a posterior distribution that is conditioned on a (finite) window of past observations 
\begin{equation}
    \label{eq:FiniteMemoryDA}
    p(\bx_k\vert \by_{1:k})\approx p(\bx_k\vert \by_{k-m:k}),
\end{equation}
where the ``memory'' $0<m<k$ is yet to be determined.
The reason is that the influence of past observations on Kalman filter updates decays exponentially with time under broad conditions \citep{ExponentiallyForgettingKFs}.
We believe it is reasonable to assume that observations in the distant past have no influence on the current state in nonlinear, chaotic dynamical systems such as Earth's atmosphere.
Using linearization and tangent linear adjoint models, we can in fact derive similar results for nonlinear problems intuitively (see Appendix~\ref{sec:NonlinExpDecay}).
One may view the approximation in~\eqref{eq:FiniteMemoryDA} as a form of ``temporal localization,'' dampening the influence of observations in the distant past on current state estimates.

The approximation~\eqref{eq:FiniteMemoryDA}, which we refer to as ``observational memory,'' is crucial for the design of diffusion DA systems, because it does away with the time-evolving, or ``cycling'' prior, characteristic of traditional DA systems.
In the context of diffusion modeling for DA, the traditional cycling paradigm implies that the diffusion model must be retrained after each DA cycle \citep{HM26,BAO2024,bao2025nonlinear,bao2024ensemble}.
To build a diffusion DA system without cycling one needs to use an observational memory, $m$.  
In terms of training data, the observational memory implies a training data set of current states, $\bx_k$, accompanied by a window of past observations $\by_{k-m:k}$.
Such a system approximates the posterior distribution $p(\bx_k\vert \by_{k-m:k})$, which has a higher forecast skill than a system without observational memory ($m=0$), which targets the posterior distribution $p(\bx_k\vert \by_k)$ \citep{HM26}.
The memory $m$ should be tuned or perhaps can be deduced from an analysis of the various time scales of the system.
Our goal here is not to provide guidelines for how to find the memory $m$, but rather to show that a \emph{finite} memory can be used to construct a diffusion DA system whose accuracy is similar to a diffusion DA system with an \emph{infinite} memory, but without the need for re-training at each DA cycle.

\subsection{Diffusion models with observational memory for DOP}
\label{sec:DiffDOPMemory}
Similar ideas apply to DOP. 
We approximate a posterior distribution with an ``infinite'' memory by a posterior distribution with a \emph{limited} memory:
\begin{equation}
    p(\by_{k+1}\vert \by_{1:k}) \approx p(\by_{k+1}\vert \by_{k-m:k}),
\end{equation}
where $0<m<k$.
The training data set for a diffusion DOP system thus consists of a (large) set of ``future'' observations $\by_{k+1}$, along with inputs $\by_{k-m:k}$. 

We note that current DOP systems are implemented exactly like this, with $m$ corresponding to a 12hr time window \cite{AIDOP,AIDOPUpdate}, recently extended to two days by \cite{AIFSDOP}.
Current DOP systems, however, do not (yet) utilize denoising diffusion models.

\subsection{Feasbility and limitations} 
Building training data sets for diffusion DA or DOP seems feasible
using ERA5 and similar atmospheric ``data,'' along with the observational record,
and one may attempt to train a large NN, using several GPUs or TPUs (as is currently underway for DOP \citep{AIDOP,AIDOPUpdate,AIFSDOP}).
We do not have access to such computational resources (few academics do), 
and we simply illustrate how to construct such diffusion DA\slash DOP systems for a canonical linear and Gaussian system.
While far from realistic, our analysis and numerical experiments demonstrate the usefulness of observational memory in diffusion DA\slash DOP systems (using several past observations) compared to diffusion DA\slash DOP systems \emph{without} observational memory (using only the current or most recent observation).
The experiments are further designed to separate discretization errors in forward\slash reverse processes from errors arising from the NN approximations of the denoiser (see Section~\ref{sec:SourcesOfErrors}) which, as far as we know, has not been studied before in this context.

\section{Case study with a canonical linear system}
\label{sec:NumericalIllustration}
Following our earlier work \citep{HM23}, we consider as a forecast model a forward Euler discretization of the SDE
\begin{equation}
    \label{cycling bickel}
    \text{d}x = -\frac{1}{2}x\text{d}t+\text{d}W,
\end{equation}
where, again, $W$ is a Wiener process and the discretization is 
\begin{equation}
    \label{eq:DiscreteModel}
    x_{k+1} = M x_k + \sqrt{\Delta t} w_k,
\end{equation}
with time step $\Delta t = 0.1$, $M=1-0.5\Delta t = 0.95$, and where $w_k\sim\mathcal{N}(0,1)$ are iid standard normal random variables.
Direct, but noisy, observations 
\begin{equation}
    \label{eq:DiscreteObs}
    y_k = x_k + \eta_k
\end{equation}
are available at each time step, where $\eta_k\sim\mathcal{N}(0,R)$, with $R=1$ throughout all experiments.
Note that we use a \emph{scalar} system, but the results trivially extend to the vector case discussed in our earlier work \citep{HM23}. 

\subsection{Exact DA with and without observational memory}
\label{sec:ExactDA}
We first consider a DA system with ``infinite'' observational memory, targeting the posterior distribution $p(x_k\vert y_{1:k})$.
This posterior distribution is Gaussian for the linear model and observation setup described above and the mean and variance of the Gaussian posterior distribution can be computed via the Kalman filter (see Section~\ref{sec:TraditionalDAReview}).
Since the model and observation equations are also time invariant, i.e., the model $M$ or the observation errors $r$ do not change with time, the forecast variance $P_\text{f}$ and the posterior covariance $P_k$ converge for large enough $k$ to steady-state values.
The steady state forecast variance $P_\text{f}^\infty$ satisfies the quadratic (Riccati) equation
\begin{equation}
    \label{eq:Riccati}
    P_\text{f}^\infty = M^2\left(P_\text{f}^\infty-\frac{(P_\text{f}^\infty)^2}{P_\text{f}^\infty + R} \right)+\Delta t,
\end{equation}
which can be solved for $P_\text{f}^\infty\approx0.32$ (matrix versions of this equation are well-known for multivariate systems).
The steady state posterior variance $P_\infty$ can then be obtained by solving
\begin{equation}
    \label{eq:Pinfty}
    P_\text{f}^\infty = M^2P_\infty +\Delta t,
\end{equation}
for $P_\infty$ to obtain $P_\infty\approx 0.24$.
Thus, a well-working DA system with infinite memory should be characterized by a mean squared error (MSE) and posterior variance of about $0.24$.

In contrast, a DA system \emph{without} memory uses the posterior distribution
\begin{equation}
    p(x_k\vert y_k) \propto p(x_k) p(y_k\vert x_k),
\end{equation}
which requires only the current observation $y_k$.
In this case, the prior $p(x_k)$ is the climatological distribution of the state $x_k$ of the dynamics~\eqref{eq:DiscreteModel}, which can be shown to be a Gaussian with mean~$\mu_\text{c}=0$ and variance
\begin{equation}
    \label{eq:VarClim}
    P_c=\frac{4}{4-\Delta t} \approx 0.98,
\end{equation}
\citep[see, e.g.,][]{HM26}.
Since $y_k\vert x_k\sim\mathcal{N}(x_k,R)$, a well-working DA system without memory should be characterized by a MSE and posterior variance
\begin{equation}
\label{eq:PostVarNoMem}
    P_\text{no mem.} =\text{var}(x_k\vert y_k)=\frac{P_c\cdot R}{P_c+R} \approx 0.49,
\end{equation}
which is about twice as large as what can be obtained with a DA system with infinite memory.
Moreover, the posterior variance defined by the climatological prior in~\eqref{eq:PostVarNoMem} is independent of $k$, but the posterior variance $P_\infty$ for a DA system with infinite memory, implicitly defined by~\eqref{eq:Riccati} and~\eqref{eq:Pinfty}, assumes that $k$ is large.

A DA system with \emph{finite} observational memory $m$ targets the posterior distribution
\begin{equation}
    p(x_k\vert y_{k-m:k})\propto p(y_k\vert x_k) p(x_k\vert y_{k-m:k-1}).
\end{equation}
We can compute the mean and variance of this posterior distribution recursively using the Kalman filter.
Specifically, at time $k$, we first use the observation $y_{k-m}$ to perform a Kalman filter update to the climatological mean ($\mu_c=0$) and the climatological variance $P_c$ in~\eqref{eq:Pinfty}.
The result is updated by a Kalman step using the observation $y_{k-m+1}$ and the process is repeated until we reach the observation $y_k$.
In essence, we ``spin-up'' a Kalman filter from the climatological mean and variance using the observations $y_{k-1:m}$.

We illustrate the effects of observational memory in DA in Figure~\ref{fig:DiffDA}(a),
where we plot MSE and posterior variance $\text{var}(x_k\vert y_{k-m})$, averaged over $10^4$ DA cycles as a function of the memory $m$.
Here, the MSE (for a scalar) is defined by 
\begin{equation}
\label{eq:DA_MSE}
    \text{MSE}_\text{DA} = (x_k^t-\mu_k^\text{DA})^2,
\end{equation}
where $x_k^t$ is the true state at time $k$ and where $\mu_k^\text{DA}$ is the mean of $x_k\vert y_{k-m:k}$.
As expected, the system without memory ($m=0$), has an MSE and variance of about 0.5 (see above).
As the observational memory $m$ increases, MSE and variance decrease and for large enough memory ($m\approx 9$), MSE and posterior variance settle on values close to 0.24, which is identical to what a DA system with infinite memory can achieve, indicating that a finite memory is sufficient \citep{ExponentiallyForgettingKFs}.
Note that the ``exact DA'' in Figure~\ref{fig:DiffDA}(a) does not use ensemble approximations, but rather computes the posterior mean and variance of $x_k\vert y_{k-m:k}$ analytically.

\subsection{Diffusion DA with and without observational memory}
\label{sec:DiffDAWithMemory}
\begin{figure}[tb]
    \centering
    \includegraphics[width=1\linewidth]{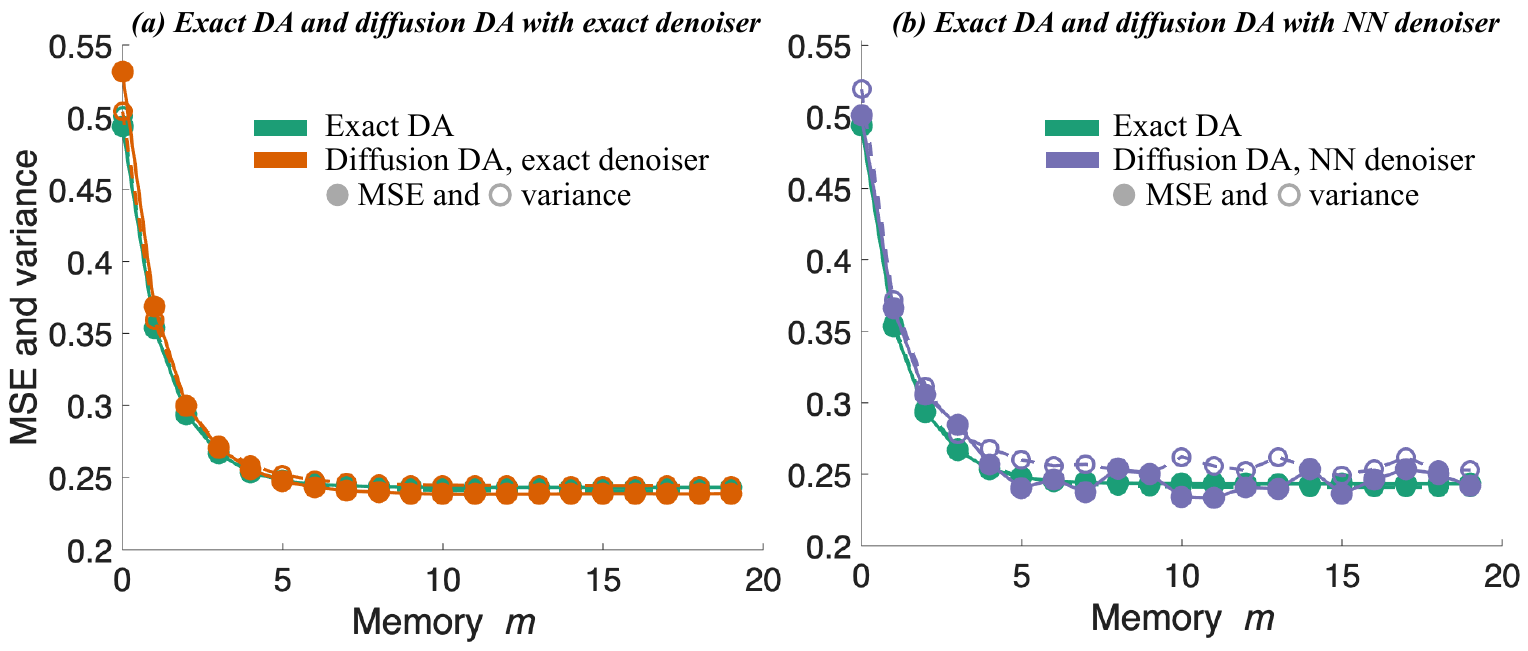}
    \caption{(a) MSE (filled circles) and posterior variance (open circles) as a function of the observational memory $m$ in a diffusion DA system.
    (a) Comparison of an exact DA (green) with a diffusion DA with an exact denoiser (orange).
    (b) Comparison of an exact DA (green) with a diffusion DA with a NN denoiser (purple).}
    \label{fig:DiffDA}
\end{figure}

We investigate if denoising diffusion generative models can emulate the ``exact'' DA systems described above.
First, we consider a diffusion DA system \emph{without} approximation errors due to NN approximations of the denoiser and use analytical formulae for the denoiser for Gaussian distributions (see Section~\ref{sec:ApproximateAndExactDenoisers}).
Specifically, we compute the mean and variance of the posterior distributions $p(x_k\vert y_{k-m:k})$ using the procedures described in Section~\ref{sec:ExactDA}, and then use the analytical formula for the denoisier in~\eqref{eq:ExactGaussianDenoiser} in the reverse process.
Results are shown in Figure~\ref{fig:DiffDA}(a), where we plot MSE and posterior variance as a function of the memory $m$. 
MSE and variance are averaged over $10^4$ DA cycles.
The diffusion DA uses $5\cdot 10^3$ samples (ensemble members) and the reverse process is discretized with a  forward Euler scheme with time step $\Delta t_\text{d}=0.1$; the final diffusion time is $T=10$.

We find that the diffusion DA system almost perfectly emulates the  ``exact'' DA system: MSE and variance are large without observational memory and gradually approach the MSE and variance of a DA system with infinite observational memory.
Indeed, MSE and variance are largely indistinguishable for diffusion DA or ``exact'' DA systems and the differences are due to the ensemble approximation of the mean and variance, and due to time discretization errors in the reverse process.
The accuracy and equivalency of the diffusion DA system is no surprise because the use of the analytical denoiser models a situation in which the diffusion DA system is perfectly trained.

In Figure~\ref{fig:DiffDA}(b) we consider a diffusion DA system with a NN denoiser, which is constructed and trained with standard methods \citep{karras2022}.
Specifically, the NN is a ``decoder'' with two layers of width 32 and 64 and we use a fully connected feed forward architecture; the activation functions are Rectified linear units (ReLU) in both layers.
We use a training set of size $10^5$ and recall that each training datum consists of a set of observations $y_{k-m:k}$ as inputs to the denoiser and the true state at time $k$ to be used in the loss function~\eqref{eq:ConditionalDenoiserLoss}.
The training (optimization) uses Adam optimization with cosine annealing (initial learning rate $10^{-1}$ and final learning rate $10^{-6}$) for 120 epochs and a batch size of 500.
For each sample, we use 100 noising steps with noise levels drawn form a lognormal distribution, $\text{Lognormal}(-1.2,1.2^2)$ (see \cite{karras2022}).
Figure~\ref{fig:DiffDA}(b) shows the analysis MSE~\eqref{eq:DA_MSE} of the ensemble mean and the posterior variance, averaged over $10^4$ DA cycles with $10^3$ samples (ensemble members) and a diffusion time $T=10$.
We note that the MSE and variance curves match up with each other and that they are also close to what an ideal (exact) DA system can achieve (shown in green in Figure~\ref{fig:DiffDA}(b) for reference).
The MSE and variance curves of the diffusion DA system with a NN denoiser, however, are more ``noisy'' than the corresponding curves of the idealized systems because the NN approximation of the denoiser induces approximation errors that are ignored in the diffusion DA system with an exact denoiser.

In summary, our idealized numerical experiment suggests that it is indeed possible to construct a diffusion DA system with a \emph{finite} observational memory that emulates best-case DA scenarios.
Diffusion DA systems with finite observational memory require no re-training at each DA cycle and therefore, may overcome a significant computational barrier.
The near optimal performance, however, relies on sufficient training of the NN denoiser, which we ensured via a large training data set ($10^5$) and thorough training for many epochs.

\subsection{Exact DOP with and without observational memory}
We now consider DOP systems and first consider an idealized system with an infinite memory that samples the pdf $p(y_{k+1}\vert y_{1:k})$.
The variance of $y_{k+1}\vert y_{1:k}$ is given in~\eqref{eq:DOPTargetPDF} and for large $k$, $\bP_k\to \bP_\infty$.
For our setup with $H=1$, $Q=\Delta t= 0.1$, $R=1$, $M=1-0.5\Delta t=0.95$ and $P_\infty\approx0.24$ (see above), we thus obtain a posterior variance of $P_\infty^\text{DOP}\approx1.32$ which is also equal to the posterior MSE
\begin{equation}
    \text{MSE}_\text{DOP} = (y_{k+1}-0.95\,\mu_k^\text{DOP})^2,
\end{equation}
where $\mu_k^\text{DOP}$ is the posterior mean at time $k$ (the mean of $y_k\vert y_{k-m:k}$).
For large $k$ this MSE and variance are optimally small.

\begin{figure}[tb]
    \centering
    \includegraphics[width=1\linewidth]{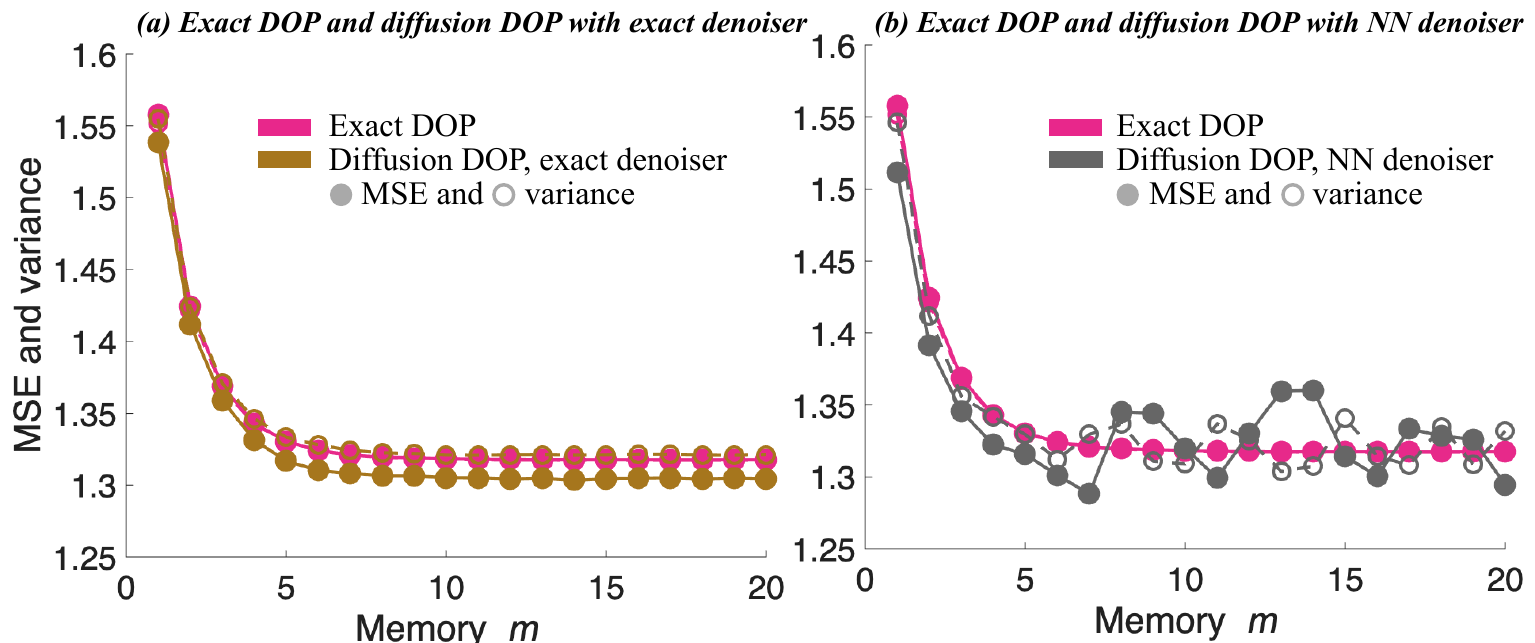}
    \caption{(a) MSE (filled circles) and posterior variance (open circles) as a function of the observational memory $m$ in a diffusion DOP system.
    (a) Comparison of an exact DOP (pink) with a diffusion DOP with an exact denoiser (brown).
    (b) Comparison of an exact DOP (pink) with a diffusion DA with a neural network denoiser (grey).}
    \label{fig:DiffDOP}
\end{figure}

An ``exact'' DOP system with memory can be designed as follows. 
We first construct $x_k\vert y_{k-m:k}$ using a DA system with memory $m$ and assimilate $m$ consecutive observations $y_{k-m:k}$ starting from the climatological prior with mean zero and variance $P_c\approx 0.98$ (as described in detail in the DA example above). 
We then use the mean and variance of $x_k\vert y_{k-m:k}$ to compute
\begin{equation}
    y_{k+1}\vert y_{k-m:k}=\bH \bM \cdot \left( \bx_{k-m:k} \right)  +\boldsymbol{\eta},
\end{equation}
(compare to~\eqref{eq:DOPTargetPDF}).
In our scalar setup, all matrices are scalar and in particular $H=1$, $M=0.95$, $Q=\Delta t$, $R=1$.
The result of using such a ``cycling'' DOP system are illustrated in Figure~\ref{fig:DiffDOP}(a),
where we plot $\text{MSE}_\text{DOP}$ and the variance of $y_{k+1}\vert y_{k-m:k}$ for various $m$, averaged over $10^4$ DOP cycles.
Note that the figure starts with a memory of $m=1$ (not zero as in the case of DA in Figure~\ref{fig:DiffDA}), because we need at least one observation to make a prediction with a DOP system.
For large $m$, we see that the MSE and variance of a DOP system with a finite memory approaches the optimal value of $P_\infty^\text{DOP} \approx 1.32$.
For $m=1$, we can also calculate the variance of $y_2\vert y_1$.
First, the posterior variance of $x_1\vert y_1$ is $(1-K)P_c$ which, with $K=P_c/(P_c+R)$, $P_c\approx 0.98$, $R=1$, is $\text{var}(x_1\vert y_1)\approx 0.49$. 
The variance of $y_2\vert y_1$ is $M^2\,\text{var}(x_1\vert y_1)_c+Q+R\approx 1.55$, as is also evident in Figure~\ref{fig:DiffDOP}.
The variance and MSE decay steadily with the memory $m$ until the optimal value is reached (at about $m\approx 8$).

We emphasize that this DOP system is not practical and should not be used as is, since it essentially requires that we use a DA system to forecast the next observation (which is not the point of DOP).
Nonetheless, this idealized DOP system illustrates that it is feasible, in principle, to design a DOP system with a finite memory that achieves virtually the same accuracy as a DOP system with an infinite memory, reiterating that $y_{k+1}\vert y_{1:k}$ is well approximated by $y_{k+1}\vert y_{k-m:k}$ for large $m$.
Finally, we note that the idealized DOP system with memory does not make ensemble approximations, but rather computes the mean and variance of $y_{k+1}\vert y_{k-m:k}$ recursively, starting from the climatological prior.
The diffusion DOP system described below does away with these limitations:
No physical model is required for both training or using the diffusion DOP;
moreover, a diffusion DOP model naturally generates an ensemble, not just a single observation\slash state estimate.

\subsection{Diffusion DOP with and without observational memory}
Next, we demonstrate that a diffusion DOP system with an ``exact'' denoiser can emulate the idealized DOP system with memory as described above. 
To set up the idealized diffusion DOP system, we compute the mean and variance of $y_{k+1}\vert y_{k-m:k}$ as described above and use the result to define the denoiser of the reverse process. 
We then perform $10^4$ DOP cycles ($5\cdot 10^3$ samples) with this idealized diffusion DOP system and plot the associated MSE and variance, averaged over all DOP cycles, in Figure~\ref{fig:DiffDOP}(a).
We note that the variances of the idealized diffusion DOP system closely follow the ``exact'' result, but the MSE is systematically a little lower than what is optimal.
Since  we can rule out approximation errors in the denoiser, we assign these differences to discretization errors in the reverse process, the finite diffusion time of $T=10$, and the ensemble approximations of the mean and variance. 
Again, we emphasize that this diffusion DOP system is not supposed to be a practical algorithm, but rather a theoretical\slash numerical demonstration that a diffusion DOP system with a sufficiently large memory $m$ can emulate a optimal DOP system with an infinite memory.

Finally, we present results of a diffusion DOP system with a NN denoiser -- such a diffusion system is, in principle, implementable and does \emph{not} require any DA calculations.
The denoiser is the same NN as described in Section~\ref{sec:DiffDAWithMemory} and we also use the same training procedure (but the training data set are now only observations, not true state and observations as in diffusion DA).
MSE and variance are shown in Figure~\ref{fig:DiffDOP}(b) for a system with $10^3$ samples, a diffusion time $T=10$ and diffusion time step $\Delta t=0.1$ (same setup as in the diffusion DA experiments). 
We note that the MSE and variance curves are more noisy than for the diffusion DOP system with an exact denoiser (Figure~\ref{fig:DiffDOP}(a)), which we also observed in the experiments with diffusion DA (Figure~\ref{fig:DiffDA}(b)).
We attribute deviations of MSE and variance of the diffusion DOP systems from the idealized case(s) to a combination of errors arising from NN approximations of the denoiser, interacting with discretization errors in the reverse process, the finite diffusion time $T=10$ and the finite training data set.
We can conclude, however, that it is possible to train diffusion DOP systems that are competitive and near optimal, even with a simple NN design and a straightforward training process. 
Most importantly, note that the diffusion DOP systems with a larger memory are more accurate than diffusion DOP systems with no or a short memory (this is true for exact system, diffusion systems with exact denoisers and diffusion systems with NN denoisers).
The results thus suggest that it is important to incorporate windows of past observations into diffusion DOP systems to achieve a near optimal accuracy.

\section{Summary and discussion}
\label{sec:SummaryAndDiscussion}
\subsection{Summary}
We consider denoising generative diffusions for data assimilation (DA) and direct observation prediction (DOP).
The goal in DA is to approximate the posterior distribution $p(x_k\vert y_{1:k})$, while DOP targets the distribution $p(y_{k+1}\vert y_{1:k})$, where $y_{1:k}$ is shorthand notation for the set of observations from time 1 to time $k$ ($y_{1:k}=\{y_1,y_2,\dots,y_k\}$) and $x_k$ is the state.
We note that both DA and DOP use conditional distributions which involve the \emph{entire} time history of observations.

Linear systems theory suggests that $x_k\vert y_{1:k}$ is well approximated by $x_k\vert y_{k-m:k}$ for sufficiently large memory $m$ because the prediction of a Kalman filter on past observations decays exponentially \citep{ExponentiallyForgettingKFs}.
Similar results hold for chaotic systems, because a chaotic system quickly ``forgets'' its initial condition (see Appendix~\ref{sec:NonlinExpDecay}).

The fact that only recent observations impact current state estimates suggests that one can design DA systems with a finite ``observational memory,'' i.e., approximating $p(x_k\vert y_{k-m:k})$, that achieve the same accuracy as a cycling DA system  that approximates $p(x_k\vert y_{1:k})$). 
Similarly, it should be feasible to design DOP systems with a finite observational memory (approximating $p(y_{k+1}\vert y_{k-m:k})$) that achieve the same accuracy as a DOP system with an infinite observational memory (approximating $p(y_{k+1}\vert y_{1:k})$).
We demonstrate that this is indeed the case with a scalar, linear system.
Using a linear system enables us to demonstrate the effects of an observational memory \emph{without} accounting for additional errors from ensemble approximations or approximation errors within diffusion models.

Distributions that are conditioned only on a window of recent observations are amenable to diffusion modeling,
while distributions that are conditioned on \emph{all} past observations are not. 
The reason is that diffusion models that are conditioned on \emph{all} past observations require re-training at every DA\slash DOP cycle and, therefore, are numerically not efficient (see also \cite{HM26,BAO2024,bao2025nonlinear,bao2024ensemble}).
Conditioning only on a set of recent observations does away with frequent retraining and, therefore, makes diffusion models more efficient for DA or DOP, while not sacrificing accuracy.

We demonstrated how to set up the training data for constructing diffusion DA and DOP systems with a finite observational memory and showcased their performance in idealized numerical experiments with a canonical, scalar linear system.
The use of the linear system enables us to properly ``ground-truth'' the diffusion models, i.e., we can check that the diffusion systems sample the desired target distributions accurately.
The numerical experiments suggest that it is indeed feasible to train diffusion DA or DOP systems on a set of recent observations and that these systems can achieve near optimal accuracy.

In the process, we also explained that diffusion models suffer from three sources of errors that interact: (i) finite training data sets; (ii) neural network approximations of the denoiser; and (iii) discretization errors in the forward and reverse processes.
We could separate the sources of error in the numerical experiments by considering ``exact'' denoisers for Gaussian distributions.
The separation of errors, enabled by focusing on a linear system, is important here.
First, we can show that in the absence of ensemble or diffusion approximation errors, DA\slash DOP systems with a finite memory can be virtually as accurate as DA\slash DOP systems with an infinite memory, which underlines the importance of incorporating a window of past observations into diffusion DA\slash DOP systems.
Second, we can construct an exact denoiser that suppresses approximation errors induced by an NN denoiser and, therefore, emulates the performance of a well-trained diffusion DA\slash DOP system. 
With an exact denoiser, a large ensemble size and appropriate discretization of the reverse process, diffusion DA\slash DOP systems operate near optimally.
Finally, using a NN denoiser on the same linear system reveals the additional errors that stem from the NN approximation of a denoiser and which can be assumed to be present in all ``real-life'' diffusion DA\slash DOP systems.

%Subsequent work should systematically investigate how discretization errors in the forward and reverse process interact with errors induced by the NN approximation of the denoiser.
%Perhaps more importantly, it will become important to understand how the training requirements of a diffusion model for DA or DOP scale with dimension. 
%In this context, we should determine how the approximation error induced by the NN approximation of the denoiser scales with the dimension of the problem and training data set.
%Methods akin to covariance localization in ensemble data assimilation may then become important for diffusion models \citep{gottwald2025localizeddiffusionmodels}.

Our results are obtained by considering a scalar linear model and one should be careful to draw conclusions about realistic or real DA and DOP systems.
As indicated above, we use the linear setup here largely because it allows us to separate sources of errors in diffusion models and to partially avoid the construction and training of neural network denoisers (which we are not experts in).
Nonetheless, our linear study is sufficient to underline the importance of an observational memory in diffusion DA or DOP system in nonlinear models: If linear models require an observational memory for improved accuracy, then it only seems natural to assume that nonlinear models will have this requirement as well.

In our study, we neglected ensemble score filters \citep{BAO2024,bao2025nonlinear,bao2024ensemble} and ``guided diffusion models'' \citep[see, e.g.,][]{manshausen2024,rozet2023,simpleObLike2024}, both of which have become popular over the past few years.
The reason is that extending our study to these methods is not straightforward and perhaps warrants additional papers.
To keep this paper concise and on point, we leave these ideas for future work. 
%Incorporating an observational memory into guided diffusions, for example, will have to include a set of past observations during guiding, which, to the best of our knowledge, has never been considered.
%

\subsection{Discussion}
At first glance, the diffusion DA systems we describe here are ``model-free,'' in the sense that the diffusion DA system consists entirely of a (discretized) reverse process and a neural-network denoiser;
no ``physics'' model, i.e., no discretization of the dynamics of the atmosphere, is needed for diffusion DA.  
The reason is that a diffusion DA system does not ``cycle,'' i.e. it does not need a model to push the posterior forward to the next set of observations.
A physical model, however, does appear during the building of the training set:
When training a diffusion model with ERA5 or on synthetic true states generated by a model, then a physical model has been used to construct the diffusion DA system.
Thus, while the deployment of a diffusion DA system is ``model-free,'' the training of a diffusion DA system requires a physics-based model and, possibly traditional DA (e.g., when using ERA5 for training).
Direct observation prediction, however \emph{never} uses a physical model, either in training or when deployed for prediction, and this technique is therefore truly ``model-free.''

Our study further suggests some broader implications for DA in general, i.e., independent of using a diffusion model to perform DA or DOP.
\begin{enumerate}

\item For more than 60 years, cycling DA, i.e., iteratively updating the last state estimate with the newest set of observations, has been \emph{the} established DA paradigm.
Our study shows that an AI-based DA system that utilizes a climatological prior in combination with an appropriately chosen observation window can perform just as effectively as a fully cycled system, indicating that strict adherence to the ``cycling paradigm'' may no longer be needed.

\item Iterative cycling over fairly short forecast windows naturally keeps the innovations small, which leads to better performance of DA techniques that require near-Gaussianity, such as the EnKF~\citep{H11}.
ML\slash AI techniques with fully nonlinear function fitting do not rely on small innovations and, therefore, allow for a deviation from the iterative cycling paradigm.

\item Reusing observations is typically frowned upon because the derivation of the cycling DA equations requires that the prior be independent of the latest set of observations. 
In climatological prior-based DA systems as described here, however, one not only \emph{can} use previously assimilated observations but --based on our results-- should reuse as many observations as possible.  
\end{enumerate}

Finally, a new AI-DOP system, called AIFS-DOP \citep{AIFSDOP}, was published by ECMWF while we were writing this paper and conducting our study.
AIFS-DOP does not use diffusion models for ensemble forecasting, but makes one forecast given a set of observations.
Nonetheless, our theory nicely aligns with what has been implemented in AIFS-DOP: The system requires a ``warm-up'' period during which the AIFS-DOP system encounters observations over the past two days.
While the AIFS-DOP paper makes no precise statements about how the two-day window was determined or what the forecasts errors for shorter windows were, we interpret the results reported by \cite{AIFSDOP} as a practical illustration of the theory we developed here, i.e., that using a window of past observations is important for the accuracy of AI-driven DA or DOP.

%%%%%%%%%%%%%%%%%%%%%%%%%%%%%%%%%%%%%%%%%%%%%%%%%%%%%%%%%%%%%%%%%%%%%
% ACKNOWLEDGMENTS
%%%%%%%%%%%%%%%%%%%%%%%%%%%%%%%%%%%%%%%%%%%%%%%%%%%%%%%%%%%%%%%%%%%%%
\section*{Acknowledgements}
MM is supported by the U.S. Office of Naval Research (ONR) Grant N000142512298.
DH  is supported by the U.S. Office of Naval Research (ONR grant N0001422WX00451.

%%%%%%%%%%%%%%%%%%%%%%%%%%%%%%%%%%%%%%%%%%%%%%%%%%%%%%%%%%%%%%%%%%%%%
% DATA AVAILABILITY STATEMENT
%%%%%%%%%%%%%%%%%%%%%%%%%%%%%%%%%%%%%%%%%%%%%%%%%%%%%%%%%%%%%%%%%%%%%
% 
%
\section*{Data statement}
The code used for making the figures will be made available on github once the paper is accepted for publication.

%%%%%%%%%%%%%%%%%%%%%%%%%%%%%%%%%%%%%%%%%%%%%%%%%%%%%%%%%%%%%%%%%%%%%
% APPENDIXES
%%%%%%%%%%%%%%%%%%%%%%%%%%%%%%%%%%%%%%%%%%%%%%%%%%%%%%%%%%%%%%%%%%%%%
%
\appendix[A]
\section{Exponential decay in nonlinear systems}
\label{sec:NonlinExpDecay}
Let the nonlinear forward model from time $t_{0}$ to $t_{n}$ be $x_{n} = \mathcal{M}(x_{0})$. The tangent linear model (TLM) evaluating perturbation growth is the Jacobian $\mathbf{M} = \frac{\partial \mathcal{M}}{\partial x_{0}}$.  In a nonlinear, chaotic system, perturbations along the unstable manifold grow exponentially. Over a sufficiently long window $\Delta t = t_{n} - t_{0}$, the dominant singular values of $\mathbf{M}$ scale as $\exp(\lambda \Delta t)$, where $\lambda > 0$ is the leading finite-time Lyapunov exponent.  

If we break the window into $n$ discrete time steps, the forecast error covariance at time $t_n$ is the sum of the evolved initial uncertainty and the evolved model error injected at each time step $i$:
\begin{equation}
    \label{A: summation prior}
    \mathbf{P}_{n} = \mathbf{M}_{0 \to n} \mathbf{P}_{0} \mathbf{M}_{0 \to n}^{T} + \sum_{i=1}^{n} \mathbf{M}_{i \to n} \mathbf{Q}_{i} \mathbf{M}_{i \to n}^{T}
\end{equation}
where $\mathbf{M}_{i \to n}$ is the tangent linear model from time $t_i$ to $t_n$. $\mathbf{Q}_{i}$ is the instantaneous model error added at step $i$.  In a chaotic system with a leading finite-time Lyapunov exponent $\lambda > 0$, the tangent linear operator  exponentially amplifies perturbations along the unstable manifold.  The first term, $\mathbf{P}_{0}$, is propagated over the full window $\Delta t = t_n - t_0$. Its contribution to $\mathbf{P}_{n}$ scales as $\exp(2\lambda \Delta t)$.  An instantaneous model error $\mathbf{Q}_{i}$ injected at a later time $t_i$ is only propagated over the remaining time $t_n - t_i$. Its contribution scales as $\exp(2\lambda(t_n - t_i))$. Because exponential growth heavily penalizes shorter propagation times, the term that has been growing the longest (which is the initial uncertainty) dominates the summation in (\ref{A: summation prior}). 

Thus, $\mathbf{P}_{n} \approx \mathbf{M}_{0 \to n} \mathbf{P}_{0} \mathbf{M}_{0 \to n}^{T}$, which means the variance at $t_{n}$ scales as $\exp(2\lambda \Delta t)$.  The influence of an observation, $y_{n}$, at time $t_{n}$ on the initial state, $x_{0}$, is determined by the gain matrix, $\mathbf{K}_{0}$, configured as a smoother. Assuming a linear observation operator, $\mathbf{H}$, and observation error covariance, $\mathbf{R}$, the gain is:
\begin{equation}
    \mathbf{K}_{0} = \mathbf{P}_{0} \mathbf{M}^{T} \mathbf{H}^{T} (\mathbf{H} \mathbf{P}_{n} \mathbf{H}^{T} + \mathbf{R})^{-1}
\end{equation}
We can evaluate the asymptotic behavior of $\mathbf{K}_{0}$ by examining its two primary components: 
\begin{enumerate}
    \item $\mathbf{P}_{0} \mathbf{M}^{T}$ scales with the adjoint $\mathbf{M}^{T}$, growing at the rate $\exp(\lambda \Delta t)$.       
    \item As forecast errors grow to swamp observation errors, $\mathbf{H} \mathbf{P}_{n} \mathbf{H}^{T} \gg \mathbf{R}$. The inverse $(\mathbf{H} \mathbf{P}_{n} \mathbf{H}^{T} + \mathbf{R})^{-1}$ therefore scales with $\mathbf{P}_{n}^{-1}$, which decays as $\exp(-2\lambda \Delta t)$.
\end{enumerate}
Multiplying the scaling factors of these two components yields the asymptotic behavior of the smoother gain:
\begin{equation}
    \label{K0}
    \mathbf{K}_{0} \propto \exp(\lambda \Delta t) \times \exp(-2\lambda \Delta t) = \exp(-\lambda \Delta t)
\end{equation}
This establishes that the sensitivity of the analysis at $t_{0}$ to an observation at $t_{n}$ decays exponentially in time, governed strictly by the system's leading Lyapunov exponent and does so in linear and nonlinear systems.  Note that the fact that the cross-covariance grew as $\exp(\lambda \Delta t)$ states that the covariance between $t=0$ and $t=t_n$ grows exponentially, which appears to imply that the covariance through time increases indefinitely.  This only happens because we are calculating the cross-covariance using a TLM.  If we were to calculate this cross-covariance with a fully-nonlinear Monte-Carlo estimation procedure ($E[(x_0-\overline{x}_0)(x_n-\overline{x}_n)]$) we would find that the cross-covariance does not grow, but actually decays.  This is because all physically realistic problems live on a finite attracting manifold and therefore this cross-covariance cannot grow forever.  For this reason, the decay in equation (\ref{K0}) is actually an upper bound and realistic nonlinear problems are likely to decay even faster.  
    
%%%%%%%%%%%%%%%%%%%%%%%%%%%%%%%%%%%%%%%%%%%%%%%%%%%%%%%%%%%%%%%%%%%%%
% REFERENCES
%%%%%%%%%%%%%%%%%%%%%%%%%%%%%%%%%%%%%%%%%%%%%%%%%%%%%%%%%%%%%%%%%%%%%
% Make your BibTeX bibliography by using these commands:
\bibliographystyle{ametsocV6}
\bibliography{references}

\end{document}